\documentstyle[preprint,floats,epsfig,aps]{revtex}
\begin{document}
\def\DESepsf(#1 width #2){\epsfxsize=#2 \epsfbox{#1}}
\input{epsf}

\begin{center}
 \vskip 15mm
{\large Is the Zee model neutrino mass matrix ruled out?}
 \vskip 15mm
Xiao-Gang He\\
Department of Physics, Nankai University, Tianjin\\
and \\
Department of Physics, National Taiwan University, Taipei
\vskip 15mm
\end{center}

\begin{abstract}
A very economic model of generating small neutrino masses is the Zee model.
This model has been studied extensively in the literature with most of the 
studies concentrated on the simplest version of the model where
all diagonal entries in the mass matrix are zero. SNO, and KamLAND data 
disfavor this simple version, but only when one also combines information 
from atmospheric and K2K data, can one rule out this model with high 
confidence level. We show that the simplest version of Zee model is ruled 
out at $3\sigma$ level. The original Zee model, however, contains more 
than enough freedom to satisfy constraints from data. We propose a
new form of mass matrix by naturalness consideration.
This new form of mass matrix predicts that $m_{\nu_3} = 0$, and
$\tan^2\theta_{solar}$ increases with $|V_{e3}|$. For the best fit value of
$\tan^2\theta_{solar}$, $|V_{e3}|$ is sizeable but below the upper bound.
\end{abstract}

\newpage

There are abundant data\cite{1,2,3,4,5,6} from solar,
atmospheric, laboratory and recent long baseline (K2K and KamLAND) 
experiments on
neutrino mass and mixing. It is certain that some of the neutrinos
have non-zero masses and also different neutrino spices mix 
with each other. In the minimal Standard Model (SM) in which there 
is just one Higgs doublet in the scalar sector and there are 
no right-handed neutrinos, neutrinos
are massless. In order to have non-zero neutrino masses and mixing, one
must go beyond the minimal SM. 

There are different possible ways to generate neutrino masses.
A very economic way of generating neutrino masses is to introduce
a charged scalar and an additional Higgs doublet into the minimal SM as 
proposed by Zee\cite{8}. The Zee model provides a 
natural mechanism to generate small neutrino masses because they 
can only be induced at loop level, and also suggests special forms 
for the mass matrix. 
If one imposes a discrete symmetry 
such that only one of the Higgs doublets couples to the leptons as
suggested by Wolfenstein\cite{9}, one obtains a simple mass matrix with 
all diagonal entries zero. We will refer this simple version as the 
Zee-Wolfenstein model.
This model has been studied extensively in the 
literature\cite{8,9,10,11,11a,12}. 
In this paper we further study the Zee model using the most recent 
experimental data. We show that the 
Zee-Wolfenstein model is ruled out at the 99.73\% ($3\sigma$) C.L.. 
However the original Zee model 
contains more than enough freedom to satisfy experimental 
constraints.
We propose a new form of neutrino mass matrix 
resulting from naturalness condition. This model predicts that 
$m_{\nu_3} = 0$, and 
$\tan^2\theta_{solar}$ increases with $|V_{e3}|$. For the best fit value of
0.4 for $\tan^2\theta_{solar}$, $|V_{e3}|$ is sizeable but below the 
3$\sigma$ upper bound.

The Zee model contains, in addition to the gauge bosons and the 
minimal fermion contents,  a singlet
scalar $h$ and two Higgs doublets $\phi_{1,2}$ transforming under the
SM gauge group $SU(3)_C\times SU(2)_L\times U(1)_Y$ as (1,1,1) and 
(1,2,-1/2). With these particles it is not possible to have tree level
neutrino masses from renormalizeable Lagrangian, but it is possible at
one loop level. The relevant terms in the Lagrangian are\cite{8},

\begin{eqnarray}
L =  - \bar l_{dR} \tilde f^{\phi,db}_{\gamma}\phi^i_{\gamma}\psi^j_{bL}\epsilon_{ij}
- \psi^{Ti}_{aL} \tilde f^{ab}C\psi^j_{bL}\epsilon_{ij} h - 
M^{\alpha\beta}\phi^i_{\alpha}\phi^j_\beta\epsilon_{ij}h + H.C.,
\end{eqnarray}
where $\psi^i_{aL} = (\nu_{aL}, e_{aL})$ and $l_{aR}$ are the 
left- and right-handed 
leptons with ``a'' the generation index and 
``i,j'' the $SU(2)_L$ indices. $\epsilon_{ij}$ is the anti-symmetric symbol.
$C$ is the Dirac charge conjugation matrix. $\tilde f^{\phi, ab}_\gamma$ 
are the Yukawa couplings responsible for charged lepton masses. 
$\tilde f^{ab}$ is an anti-symmetric matrix in generation
indices $a$ and $b$ due to Fermi statistics.

The mass matrix $\tilde m$ for 
the charged leptons is given by, $\tilde m = 
(v_1 \tilde f^{\phi}_1 + v_2 \tilde f^{\phi}_2)
=v(\sin\beta \tilde f_1^{\phi} + \cos\beta \tilde f_2^\phi)$. 
Here $v_\gamma = <\phi_\gamma>$ 
are the vacuum expectation values (VEV),
$v=\sqrt{v_1^2+v_2^2} = 174$ GeV and $\tan\beta = v_1/v_2$.
The mass matrix $\tilde m$ can be
diagonalized to obtain the eigen-mass matrix 
$m = Diag(m_e,\;m_\mu,\;m_\tau)$ 
by a bi-unitary transformation multiplying two unitary matrices $V_{L,R}$
from left and right, $m = V_R \tilde m V_L^\dagger$. 

The linear combination $\phi^-_W = \cos\beta \phi_1^- 
+\sin\beta \phi^-_2$ is ``eaten'' by $W^-$. 
The physical combination which mixes with $h$ is
$\phi^- = \sin\beta \phi^-_1 - \cos\beta \phi^-_2$. We indicate 
the two mass eigenstates of masses $M_1$ and $M_2$
for the charged scalars as $h_1 = \cos\theta_Z h -\sin\theta_Z \phi^+ $ 
and $h_2 = \sin\theta_Z h + \cos\theta_Z \phi^+$. 
Here $\sin \theta_Z$ is proportional to $M_{\alpha\beta}$ 
characterizing the strength of the $h-\phi^+$ mixing. 

The terms responsible for neutrino mass generation in the previous equation, 
in the mass eigenstates basis for the charged lepton and 
scalar fields, can be written as

\begin{eqnarray}
L &=&- \bar E_R  m E_L - \bar E_R ({1\over v\tan\beta }  m
-{1\over \sin\beta}  f^{\phi}_2)\nu_L(\sin\theta_Z h^\dagger_1 - 
\cos\theta_Z h^\dagger_2)\nonumber\\
&&- 2 \nu^T_L  f C E_L (\cos\theta_Z h_1 + \sin\theta_Z h_2)
+ ...
\end{eqnarray}
where $ f^\phi_\gamma = (f^{\phi,ab}_\gamma)= V_R \tilde 
f^\phi_\gamma V_L^\dagger$, $ f = (f^{ab}) = V^*_L \tilde f V_L^\dagger$,
$E_{L,R} = (e,\; \mu,\;\tau)_{L,R}$, and $\nu_L = (\nu_1, \nu_2, \nu_3)_L$.

Exchange of charged scalars $h_{1,2}$ and charged leptons at one loop level, 
Majorana
neutrino mass term 
$L_m = (1/2)\nu^T_L M_\nu C \nu_L$ can be generated with

\begin{eqnarray}
M_\nu = A[( f  m^2 +  m^2  f^T)
- {v\over \cos\beta} ( f  m  f^\phi_2 +
 f^{\phi T}_2  m  f^T)],
\label{mass}
\end{eqnarray}
where $A = \sin(2\theta_Z) log(M^2_2/M^2_1)/(16\pi^2 v \tan\beta)$ which is of order
$O(10^{-5})$ if the $\sin(2\theta_Z)$ and 
$\tan\beta$ are both of order one. This is the general mass matrix in the Zee model\cite{12}. The mixing matrix 
is the unitary matrix $V$ which diagonalizes the mass matrix and is defined by,
$D = V^T M_\nu V$, with $D = diag(m_{\nu_1}, m_{\nu_2}, m_{\nu_3})$. 

The present experimental data on neutrino masses and mixing angles
can be summarized as follow\cite{14,14a}.
The $3\sigma$ allowed ranges for the mass-squared
differences are constrained to be: $1.6\times 10^{-3}$ eV$^2$
$\leq |\Delta m^2_{atm}| \leq 4.8\times 10^{-3}$ eV$^2$, and
$4.7\times 10^{-5}$ eV$^2$ $\leq \Delta m^2_{solar} \leq
1.7 \times 10^{-4}$ eV$^2$, with the best fit values
given by $|\Delta m^2_{atm}| = 2.5\times 10^{-3}$ eV$^2$, and
$\Delta m^2_{solar} = 7.0\times 10^{-5}$ eV$^2$.
The mixing angles are in the ranges of $0.3\leq \sin^2\theta_{atm}
\leq 0.7$ and $0.29 \leq \tan^2\theta_{solar} \leq 0.63$. 
Also the CHOOZ experiment\cite{4} gives an upper bound of $0.22$ on 
the $\nu_e -\nu_x$ (where $\nu_x$ can be either 
$\nu_\mu$ or $\nu_\tau$ or a 
linear combination) oscillation parameter for 
$\Delta m^2_{x1}=|m_x|^2 - |m_{\nu_1}|^2 > 10^{-3}$ eV$^2$. 

In the model discussed here the atmospheric neutrino and K2K data can 
be explained by oscillation between the muon
and the tauon neutrinos, and the solar neutrino and KamLAND
data explained by oscillation between the electron and muon 
(or a linear combination of
muon and tauon neutrino) neutrinos.
In this case the CHOOZ limit applies to the oscillation between the
electron and tauon neutrinos
\footnote{There are additional evidences for
oscillation between electron and muon neutrinos
from LSND experiment\cite{lsnd}. If confirmed more
neutrinos are needed to explain all the data.}.

Setting $f^{\phi}_2$ in eq. (\ref{mass}) to zero, one obtains the
famous Zee-Wolfenstein mass matrix,

\begin{eqnarray}
M_\nu = \left ( \begin{array}{rrr}
0& \tilde a &\tilde b\\
\tilde a & 0&\tilde c\\
\tilde b & \tilde c &0
\end{array}
\right ),
\label{zee}
\end{eqnarray}
where $\tilde a = A f^{e\mu}(m_\mu^2 - m_e^2)$, 
$\tilde b = Af^{e\tau}(m^2_\tau - m^2_e)$
and $\tilde c = Af^{\mu\tau}(m^2_\tau - m^2_\mu)$.
One can redefine the neutrino and charged lepton
phases such that all $\tilde a$, $\tilde b$ and $\tilde c$ 
are real. 

Unfortunately the Zee-Wolfenstein model
is now ruled out by experimental data. 
This can be seen from the following.

The above mass matrix satisfies the ``zero sum''
condition $m_{\nu_1} + m_{\nu_2} + m_{\nu_3} =0$, therefore all
the neutrino masses are determined in terms of the
mass-squared differences\cite{18}.
We have\cite{18}

\begin{eqnarray}
m_{\nu_1}^2 = -{1\over 3}\left [2 \Delta m^2_{21} +  \Delta m^2_{32}
- 2\sqrt{(\Delta m^2_{32})^2 + \Delta m^2_{21} \Delta m^2_{32} +
(\Delta m^2_{21})^2}
\right ].
\end{eqnarray}
The other two masses are given by, 
$m^2_{\nu_2} = \Delta m^2_{21}+m^2_{\nu_1}$ and $m^2_{\nu_3}
= \Delta m^2_{32} + m^2_{\nu_2}$.
The ``zero sum'' condition admits two types of mass hierarchy if 
the absolute value of $r = \Delta m_{21}^2/\Delta m_{32}^2$ 
is much smaller than one (experimentally $|r| < 0.106$ at $3\sigma$ level), 
with one of
them the normal one: $m_{\nu_3} > m_{\nu_2} > m_{\nu_1}$ and $m_{\nu_1} 
\approx m_{\nu_2}$, and
another the inverted one: $ |m_{\nu_2}| > |m_{\nu_1}| > |m_{\nu_3}|$ and 
$m_{\nu_2} \approx -m_{\nu_1}$.
One finds that $|x| = |m_{\nu_1}/m_{\nu_2}|$ is determined to be 
very close to one.

The mass matrix element $M_{11} = 0$  
leads to, $V^2_{e1}m_{\nu_1} + V_{e2}^2 m_{\nu_2} + V_{e3}^2 m_{\nu_3} = 0$, 
which can 
be rewritten as 

\begin{eqnarray}
V_{e2}^2 = {-x + (1+2x) V_{e3}^2 \over 1 - x}.
\end{eqnarray}
Since 
$|x|$ is smaller but close to one, the above equation only allows
negative $x$ for small $V_{e3}^2$ implying that only the
inverted mass hierarchy is possible.
One thus obtains a minimal $V_{e2,min}^2$ of 
$V_{e2}^2$ close to $(1 - V_{e3,max}^2)/2\approx
0.47$, while data from SNO and KamLAND prefers a smaller $V_{e2}^2$.
Therefore SNO and KamLAND data disfavor the Zee-Wolfenstein model.
This has been noticed in Ref.\cite{11a}. However, we would like to
point out that although the Zee-Wolfenstein 
model can not produce the central values for the mixing and mass difference
from solar and KamLAND data, the present data can not rule out the model 
at more than even 2$\sigma$ level.

To have a more quantitative
statement, we have carried out a detailed study and the results are 
shown in Figure 1.
The dashed lines in Figure 1 are for 
$\tan^2\theta_{solar}$ 
($\sin^22\theta_{solar}
= 4|V_{e1}|^2|V_{e2}|^2$) with two values (0.22 and 0.15) 
of $V_{e3}$ as a function of r. 
When $|V_{e3}|$ decreases, $\tan^2\theta_{solar}$
increases. 
$\tan^2\theta_{solar}$
is about 0.53 for the $3\sigma$ upper bound of $|V_{e3}|$,
and becomes larger than the $3\sigma$ allowed
value of 0.63 when $|V_{e3}|$ decreases to be lower than 0.15.
One therefore can take $|V_{e3}|$ to be larger than 0.15 at 3$\sigma$ level. 
It is clear that the model is not possible to produce 
the best fit value of 0.4 for 
$\tan^2\theta_{solar}$. However at 2$\sigma$, $\tan^2\theta_{solar}$ can
be as large as 0.54\cite{14a}.  
Therefore it is not possible to rule out the model at 
more than 2$\sigma$ level from data on solar and KamLAND.

\begin{figure}[htb]
\centerline{ \DESepsf(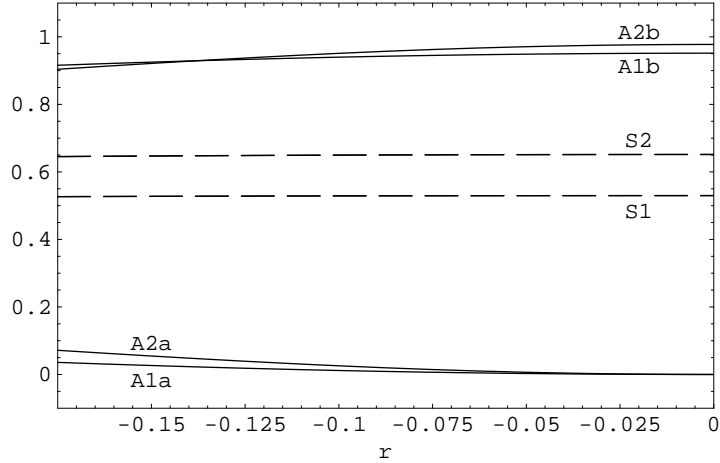 width 10cm)}\label{contour}
\smallskip
\caption{The dashed lines $S1$ and $S2$ are for $\tan^2\theta_{solar}$ as 
functions of r. 
The two solutions for $\sin^2\theta_{atm}$
are indicated by solid lines $A1a$ and $A2a$, and 
$A1b$ and $A2b$, respectively. Here the indices ``1'' and ``2'' indicate the
cases with $|V_{e3}|$ equals to 0.22 and 0.15, respectively.} 
\end{figure}

Data on $\sin^2\theta_{atm}$ can provide further constraints on the model.
The condition $M_{22} = m_1 V_{\mu1}^2
+m_2V_{\mu2}^2 + m_3V_{\mu 3}^2 = 0$ in the model can be used to determine 
$\sin^2\theta_{atm} = V^2_{\mu 3}$.
The mixing matrix $V$ can be parameterized using three 
rotation angles, for example\cite{1} $V_{e2} = s_{12}c_{13}$,
$V_{e3} = s_{13}$ and $V_{\mu 3} = s_{23} c_{13}$. 
Here $s_{ij} = \sin\theta_{ij}$ and $c_{ij}
=\cos\theta_{ij}$.
Two of the angles, $\theta_{12,13}$ can be
determined in terms of $V_{e3}$ and $r$ from previous discussions. 
The condition

\begin{eqnarray}
M_{22} = m_{\nu_1} (s_{12} + c_{12}s_{13} t_{23})^2
+ m_{\nu_2}(c_{12}-s_{12}s_{13} t_{23})^2
+ m_{\nu_3} c^2_{13}t^2_{23} = 0,
\end{eqnarray}
then determines $\theta_{23}$ in terms of $V_{e3}$ and $r$. 
Here $t_{23}=s_{23}/c_{23} = \tan\theta_{23}$.
There are two solutions for $\tan\theta_{23}$ for given 
$V_{e3}$ and y which we indicate by ``a'' and ``b''. 

In Figure 1 the solid lines show 
$\sin^2\theta_{atm}$ as a function of $r$ for $|V_{e3}|$
equals to its 3$\sigma$ allowed upper value of 0.22 and the 
allowed lower value of 0.15. From the figure 
we see that $\sin^2\theta_{atm}$ decreases for solution ``a'', and
$\sin^2\theta_{atm}$ increases for solution ``b'' as $r$ increases from
the 3$\sigma$ lower bound of -0.106 to the allowed upper bound of 0.
All solutions for $\sin^2\theta_{atm}$ are outside the 3$\sigma$
allowed range of $0.3 \sim 0.7$. For $|V_{e3}|$ smaller than 
0.15, it is possible for $\sin^2\theta_{atm}$ of 
solution ``b'' to become smaller than the 3$\sigma$
allowed upper bound.
However $|V_{e3}|$ smaller than 0.15 will drive 
$\tan^2\theta_{solar}$ to move out the $3\sigma$ allowed range. Therefore 
the combined
neutrino data on $\tan^2\theta_{solar}$ and $\sin^2\theta_{atm}$ 
rule out the Zee-Wolfenstein model at more $3\sigma$ level.

The above discussions 
show clearly that the Zee-Wolfenstein neutrino mass matrix is 
in trouble. 
That does not, however, mean that the Zee model itself 
is in trouble.
The mass matrix given in eq. (\ref{mass})
contains more than enough freedom to fit data. 
Here we encounter 
a common problem for physics beyond the SM that there are too many new 
parameters. 
Additional theoretical considerations have to be applied to narrow down
the parameters.

We find that an interesting neutrino mass matrix emerges
if one requires that there should be no large hierarchies among the
new couplings, that is all $f^{ij}$ and $f^{\phi,ab}_2$ are of the same order of
magnitude, respectively. This can be considered as a naturalness requirement. 
From eq. (\ref{mass}) one sees that all terms in 
the mass matrix are either proportional to $m_l$ or $m^2_l$. Since
$m_\tau >> m_{\mu, e}$, the leading contributions to the neutrino mass matrix 
are proportional to
$f^{ij} m^2_\tau$ and $f^{\phi,ab}_2 m_\tau$. To this order we have 

\begin{eqnarray}
M_{11}& = &-2A {v\over \cos\beta} f^{e\tau} f^{\phi, \tau e}_2 m_\tau,
\;\;M_{22} = -2{v\over \cos\beta} f^{\mu\tau}f^{\phi,\tau \mu}_2 m_\tau,\;\;
M_{33} = 0,\nonumber\\
M_{12} & = & - {v\over \cos\beta}A (f^{e\tau}f^{\phi,\tau \mu}_2
+f^{\mu \tau}f^{\phi,\tau e}_2)m_\tau,\nonumber\\
M_{13} & = & A 
f^{e\tau} m_\tau( m_\tau - {v\over \cos\beta} f^{\phi,\tau\tau}_2),
\;\;M_{23} = Af^{\mu\tau} m_\tau(m_\tau -{v\over \cos\beta} 
f^{\phi,\tau\tau}_2).
\end{eqnarray}

Without loss of generality, by appropriate choices of neutrino filed phases,
the 11, 13, 23 entries can be made real with just one physical phase 
$\delta$ in the mass matrix. One 
can rewrite the above mass matrix as

\begin{eqnarray}
M_\nu = a \left ( \begin{array}{ccc}
1&(ye^{i\delta} + x)/2&z\\
(ye^{i\delta} + x)/2&xye^{i\delta}& xz\\
z&xz&0
\end{array}
\right ),
\end{eqnarray}
with $a = |M_{11}|$, $x=|f^{\mu\tau}|/|f^{e\tau}|$, $y = |M_{22}|/xa$, $z = 
|M_{13}|/a$.

This is a highly constrained form of mass matrix. This matrix is
rank two implying that one of the neutrinos has zero mass. The
non-zero eigenvalues are given by

\begin{eqnarray}
m^2_{\pm} = {a^2\over 4}
(\sqrt{1+2 x y\cos\delta + x^2+y^2} \pm \sqrt{(1+x^2)(1+y^2+4z^2)})^2.
\end{eqnarray}

Since experimentally $\Delta m^2_{21} >0$, there are two types of eigen-mass
hierarchies, a) $m_{\nu_1} =0$, $|m_{\nu_2}| = \sqrt{\Delta m^2_{21}}= m_{-}$,
$|m_{\nu_3}| = \sqrt{\Delta m^2_{32} - \Delta m^2_{21}} = m_{+}$; and b)
$|m_{\nu_1}| = \sqrt{|\Delta m^2_{32}|-\Delta m^2_{21}} = m_{-}$,
$|m_{\nu_2}| = \sqrt{|\Delta m^2_{32}|}= m_{+}$, $m_{\nu_3} = 0$.
The five parameters
in the mass matrix are severely constrained from data on
$\Delta m^2_{21,32}$, $V_{e2}$, $V_{e3}$ and $V_{\mu 3}$. 

To have some idea about what parameter space may satisfy experimental 
constraints, let us discuss the situation with the phase $\delta$ set 
to be zero for simplicity. 
For type a) of mass hierarchy since $|r|= |\Delta m^2_{21}/\Delta m^2_{32}|$ 
is much smaller than 1, one would have $(1+xy)^2$ to be almost equal to
$(1+x^2)(1+y^2+4z^2)$. To satisfy this, $x$ should be close to $y$ and
$z$ to be much smaller than 1.  Expanding the mixing matrix elements around
$x=y$ and small $z$, we find that $(V_{e2},\;V_{\mu 2},\;V_{\tau 2})$
to be proportional to $(z,\; xz,\;-(1+x^2))$. Since $z$ is much smaller than
1, one would obtain too small a $V_{e2}$ in contradiction with solar and 
KamLAND data. This qualitative feature is not changed even if a non-zero
$\delta$ is introduced. There is no solution for the normal mass
hierarchy of type a).  

For type b) of mass hierarchy, one has $(V_{e 3}\;V_{\mu 3},\;V_{\tau 3})$ 
is proportional to $(-2x z, \;2z, x-y)$. A small $|r|$ requires $xy$ to be
close to -1. Then small $V_{e3}$, and large $|V_{\mu 3}|$ and $|V_{\tau 3}|$ 
require $x$ to be small and $2xz$ to be of order one. We indeed find solutions
for the mixing matrix satisfying experimental constraints. 
We also find that the size of $V_{e3}$ anti-correlates with 
$\tan^2\theta_{solar}$ strongly, 
that is, when $|V_{e3}|$ decreases, $\tan^2\theta_{solar}$ increases. 
If $\tan^2\theta_{solar}$ is close to its best fit value of 0.4, $|V_{e3}|$
is close to, but below, the 3$\sigma$ upper bound of 0.22.
In the following we present a sample solution with $\Delta m^2_{21,32}$ 
have their best fit values,

\begin{eqnarray}
&&m_{\nu_1} = 4.93\times 10^{-2} \mbox{eV},\;\;m_{\nu_2} = 
-5.00\times 10^{-2} \mbox{eV}, \;\;m_{\nu_3} = 0.
\nonumber\\
&&V = \left (
\begin{array}{rrr}
0.8244&-0.5312&-0.1953\\
0.2961&0.6989&-0.6511\\
0.4823&0.4789&0.7335
\end{array}
\right ).
\label{comp}
\end{eqnarray}
The $\tan^2\theta_{solar}$ is 0.415 close to the best fit value.
The value $-0.1953$ for $V_{e3}$ is below, but close to the 3$\sigma$
allowed upper bound.

In the above solution, the input parameters are: $x=-0.3$, $y=3.455$, 
$z=1.667$, 
$a=1.94\times 10^{-2}$ eV. One can choose different signs for the 
parameters $x$, $y$ and $z$.
As long as the signs for $x$ and $y$ are simultaneously changed, the
magnitudes of the eigen-masses and the mixing matrix elements are not changed.
We will stick to the signs with $x$ to be negative, $y$ and $z$ to be positive
in our later discussions.

One can also find solutions with smaller $|V_{e3}|$. For example,
with $x= -0.165$, $y=6.531$ and $z=3.030$, 
we obtain $V_{e3} = -0.11$, but $\tan^2\theta_{solar} =0.624$
which is close to the 3$\sigma$ upper bound. 

We searched for other solutions. We find that it is also 
possible to have solutions with non-zero CP violating phase $\delta$. 
For example with 
$a = 1.92\times 10^{-2} \mbox{eV}$, $x = -0.276$, $ye^{i\delta} 
= 3.467 - i 0.0573$, and $z= 1.571$, we have

\begin{eqnarray}
&&m_{\nu_1} = 4.93\times 10^{-2} e^{-i16.9^\circ} \mbox{eV},\;\;m_{\nu_2} = 
-5.00\times 10^{-2} e^{i11^\circ}\mbox{eV}, \;\;m_{\nu_3} = 0.
\nonumber\\
&&V = \left (
\begin{array}{rrr}
0.8147&-0.5048 - i 0.2311&-0.1676 - i 0.0024\\
0.3110 - i 0.1995&0.7035&-0.6071-i0.0087\\
0.4166-i 0.1619&0.4402 - i 0.0561&0.7767
\end{array}
\right ).
\label{comp}
\end{eqnarray}

The value for
$\tan^2\theta_{solar}$ is about 0.464 which is within the 1$\sigma$ region. 
The value $|V_{e3}| = 0.168$ is below the $3\sigma$ upper bound, 
but not far below. 
The CP violating Jarlskog parameter 
$J = Im( V_{11}V_{22} V_{12}^*V_{21}^*)$ is predicted to be $-0.0165$ which
may be studied in future neutrino factories. 
We have kept masses in the form with phases to illustrate the existence of 
Majorana phases which can be rotated away by multiplying 
a phase matrix from the right on $V$ obtained above.

The neutrino masses obtained in the model are in the interesting ranges. 
The sum of the absolute neutrino masses, $m_{sum} 
= |m_{\nu_1}|+ |m_{\nu_2}| + |m_{\nu_3}|$, in this model is 
around $0.1$ eV which is 
several times smaller than the recent bound of 0.69 eV from WMAP\cite{22}, 
but can be probed in 
the near future by the PLANK experiment where the 
sensitivity on $m_{sum}$ can be as low as 0.03 eV.
Laboratory neutrino mass experiments can also test the model.
A non-zero value $a =|m_{ee}|$ can 
induce neutrinoless double beta decays. 
$|m_{ee}|$ obtained here is about $0.02$ eV 
which is safely below the present bound\cite{1,24} of 0.4 eV. 
However it can be probed by future experiments, such as GENIUS, MOON and 
CUORE, where
sensitivity of about 0.01 eV may be reached.
The effective mass 
$<m_e> = \sqrt{|m_{\nu_1}V_{e1}|^2 + |m_{\nu_2}V_{e2}|^2 
+ |m_{\nu_3}V_{e3}|^2}$ measured by the end point spectrum of beta decay 
in our case is around $\sim 0.05$ eV which is unfortunately a factor 
of 2 smaller than the 
sensitivity of future KATRIN 
experiment.

The off-diagonal entries of the 
couplings $f^{ab}$ and $ f^{\phi, ab}_2$ can induce flavor changing 
interactions. One should make sure that
constraints on related parameters will not rule out the regions of the 
parameters to produce the mass matrix discussed above.
It is not possible to completely determine the couplings using just 
information from neutrino masses and mixing. We therefore 
take a simple situation 
with $f^{\phi, \tau\tau}_2 = 0$ for
illustration. 
In this case for the example given in eq. (\ref{comp}):
$f^{\phi, \tau e}_2/\cos\beta = -0.33\times 10^{-2}$,
$f^{\phi, \tau \mu}_2/\cos\beta = -(1.14 - i 0.02)\times 10^{-2}$,
$Af^{e\tau} = 0.93\times 10^{-11} (\mbox{GeV}^{-1})$, and
$Af^{\mu \tau} = -0.26\times 10^{-11} (\mbox{GeV}^{-1})$. It is interesting 
to note that 
the solution obtained here is consistent with the naturalness 
requirement that
$f^{\phi, \tau e}_2$ to be the same order of magnitude as 
$f^{\phi, \tau \mu}_2$, and $f^{e\tau}$ to be the same order of magnitude 
as $f^{\mu\tau}$. If one chooses a smaller $x$ one would obtain bigger 
hierarchy for the parameters, $f^{\mu\tau}$ and $f^{e\tau}$.
The qualitative features will not change when other values 
for the parameters are used. 

Exchange of the neutral Higgs boson $\phi_1$ (with mass $M_0$)  
can induce at tree level $l_i \to l_j l_k \bar l_k$ decays.
For the values of $f^{\phi, \tau\mu}_2$ and $f^{\phi,\tau e}_2$
obtained in the example of eq.(\ref{comp}) we have

\begin{eqnarray}
&&B(\tau \to \mu \mu \bar \mu, \mu e \bar e) \approx 3.5 \times 10^{-9}B_\tau,
\;0.80\times 10^{-13} B_\tau;\;B(\tau \to \mu \gamma)
\approx 0.76 \times 10^{-8} B_\tau; \nonumber\\
&&B(\tau \to e \mu \bar \mu, e e \bar e) \approx 2.9\times 10^{-10} B_\tau,
\;0.67 \times 10^{-14}
B_\tau;\;B(\tau \to e \gamma) \approx 0.63 \times 10^{-9} B_\tau.\nonumber
\end{eqnarray}
In the above $B_\tau =(100 (\mbox{GeV})/ M_0\tan\beta)^4 B^{SM}(\tau
\to \nu_\tau \mu \bar\nu_\mu)$ with $B^{SM}(\tau
\to \nu_\tau \mu \bar \nu_\mu )\approx 17\%$.

There are experimental constraints on the above decays with 
the 90\% C.L. bounds given by\cite{1}:
$B(\tau \to \mu \mu\bar \mu, \mu e \bar e) = 1.9\times 10^{-6}$, 
$1.7\times 10^{-6}$, 
$B(\tau \to e \mu\bar \mu, e e\bar e) = 1.8\times 10^{-6}$, 
$2.9\times 10^{-6}$, 
$B(\tau \to \mu \gamma, e\gamma) = 1.1\times 10^{-6}$, 
$2.7\times 10^{-6}$. 
For $\tan\beta$ of order one, all the branching ratios predicted 
above are safely below the experimental 
values if the mass $M_0$ is of order 
100 GeV.

Non-zero $f^{ij}$ can also induce radiative charged lepton decays by exchanging 
charged scalars.
If the parameter
$A$ is not too much smaller than a natural value of $A = 10^{-5}$ (GeV$^{-1}$), 
their contributions for the rare decays mentioned will be much smaller. The 
rare decays of the types discussed in the above will not provide significant 
constraints.

From the above discussions we see that the new form of mass matrix proposed
is consistent with present experimental data. It also predicts $m_{\nu_3} = 0$ 
and a sizeable
$|V_{e3}|$. In particular, if the error on $\tan^2\theta_{solar}$ 
is reduced and the present best fit value holds, $|V_{e3}|$ will be close to 
the 3$\sigma$ allowed upper bound. The model can be tested in the future.

\acknowledgments
I thank A. Zee for many useful suggestions.
This work was supported in part by
NSC under grant number NSC 91-2112-M-002-42,
and by the MOE Academic Excellence Project 89-N-FA01-1-4-3 of Taiwan,
\\
\\

\tighten


\end{document}